\documentclass[draftclsnofoot, onecolumn]{IEEEtran}
\IEEEoverridecommandlockouts
\usepackage{cite}
\usepackage{amsmath,amssymb,amsfonts}
\usepackage{algorithmic}
\usepackage{graphicx}
\usepackage{textcomp}
\usepackage{xcolor}
\usepackage{tabularx}
\usepackage{datatool}
\usepackage{float}
\usepackage{booktabs}

\def\BibTeX{{\rm B\kern-.05em{\sc i\kern-.025em b}\kern-.08em
    T\kern-.1667em\lower.7ex\hbox{E}\kern-.125emX}}

\begin{document}


\thispagestyle{empty}

\newcolumntype{L}[1]{>{\raggedright\arraybackslash}p{#1}}
\newcolumntype{C}[1]{>{\centering\arraybackslash}p{#1}}
\newcolumntype{R}[1]{>{\raggedleft\arraybackslash}p{#1}}

\clearpage
\pagenumbering{arabic} 

\title{Adapting to the AI Disruption: Reshaping the IT Landscape and Educational Paradigms}

\makeatletter
\newcommand{\linebreakand}{
  \end{@IEEEauthorhalign}
  \hfill\mbox{}\par
  \mbox{}\hfill\begin{@IEEEauthorhalign}
}

\makeatother
\author{
    Murat Ozer\textsuperscript{1}, 
    Yasin Kose\textsuperscript{2}, 
    Goksel Kucukkaya\textsuperscript{1}, 
    Assel Mukasheva\textsuperscript{3}, 
    Kazim Ciris\textsuperscript{1} \\[1em]
    \textsuperscript{1}\textit{School of Information Technology, University of Cincinnati, Cincinnati, Ohio, USA} \\
    \textsuperscript{2}\textit{Cybercrime and Forensic Computing, Friedrich-Alexander-Universität, Erlangen, Germany} \\
    \textsuperscript{3}\textit{Information Systems, Kazakh-British Technical University, Almaty, Kazakhstan} \\[1em]
    m.ozer@uc.edu, yasin.koese@fau.de, kucukkgl@ucmail.uc.edu, a.mukasheva@kbtu.kz, ciriskm@ucmail.uc.edu
}

\maketitle

\thispagestyle{plain}
\pagestyle{plain}

\begin{abstract}

Artificial intelligence (AI) signals the beginning of a revolutionary period where technological advancement and social change interact to completely reshape economies, work paradigms, and industries worldwide. This essay addresses the opportunities and problems brought about by the AI-driven economy as it examines the effects of AI disruption on the IT sector and information technology education. By comparing the current AI revolution to previous industrial revolutions, we investigate the significant effects of AI technologies on workforce dynamics, employment, and organizational procedures. Human-centered design principles and ethical considerations become crucial requirements for the responsible development and implementation of AI systems in the face of the field's rapid advancements. IT education programs must change to meet the changing demands of the AI era and give students the skills and competencies they need to succeed in a digital world that is changing quickly. In light of AI-driven automation, we also examine the possible advantages and difficulties of moving to a shorter workweek, emphasizing chances to improve worker productivity, well-being, and work-life balance. We can build a more incslusive and sustainable future for the IT industry and beyond, enhancing human capabilities, advancing collective well-being, and fostering a society where AI serves as a force for good by embracing the opportunities presented by AI while proactively addressing its challenges.
\end{abstract}
 
\section{Introduction}

There are turning points in human history that permanently change the course of civilization and the way we work, live, and engage with the outside world. Every technological advance, from the development of the abacus to the onset of the computer era, has opened up new avenues for creativity and increased the limits of human understanding. A new narrative, however, is beginning to emerge amid the many tales of technological wonders and human ingenuity; this narrative speaks to the dawn of a truly transformative era: the era of artificial intelligence (AI).

We have all heard the myths and stories, handed down through the ages, about how the invention of the calculator revolutionized mathematics by relieving people of the mental strain of doing calculations by hand and igniting a wave of scientific inquiry. On the other hand, the development of the computer promised unparalleled computational power, allowing individuals and organizations to tackle challenging problems at a pace and efficiency never seen before.

But this is also the point at which the story takes an unexpected turn that fundamentally and significantly distinguishes artificial intelligence from its predecessors. You see, computers and calculators still require human operators to enter data, execute commands, and interpret results. In contrast, artificial intelligence operates at an entirely different level of autonomy. Unlike its predecessors, AI transcends rather than just augments human capabilities \cite{brynjolfsson2022turing}. It performs each of these functions without the use of labor or the tangible infrastructure associated with conventional technology. It can make decisions, identify trends, and analyze big datasets.

The invention of computers, for instance, did not lead to widespread unemployment as initially feared. Instead, it opened up new vistas of employment opportunities, from software development to IT support. However, the advent of artificial intelligence represents a paradigm shift of a different magnitude altogether. While computers extended human capabilities, AI has the potential to replace them. It's not just about automating manual tasks anymore; it's about machines being able to learn, adapt, and even innovate independently. This level of autonomy has far-reaching implications for industries, economies, and societies at large. As AI continues to advance, it's not just blue-collar jobs that are at risk; white-collar professions like accounting, legal services, and even medicine are facing the prospect of automation.

Essentially, the development of artificial intelligence signals the beginning of a new chapter in human history, one in which the definition of productivity and labor itself are being altered. It is about radically changing how we view and engage with technology, not just about streamlining procedures or improving current workflows. It is crucial that we foresee the consequences of this brave new world of AI-driven automation and proactively shape laws and procedures to guarantee a time when people and machines can live in harmony. In essence, artificial intelligence (AI) represents a paradigm shift—a venture into the unknown where machines are endowed with sentience and agency. It makes it possible for algorithms to navigate challenging situations, resolve challenging issues, and even develop and adapt over time \cite{huang2021connotation}. The thought of this technological wonder's ability to completely change every aspect of our lives—from healthcare and transportation to finance and entertainment—is both amazing and unsettling.
The question of what this means for the future of work and the millions of people whose livelihoods depend on jobs that could soon be automated away lingers, though, amidst the excitement and anticipation surrounding the rise of AI. It's a question that requires careful thought, strategic planning, and a dedication to making sure the advantages of AI are distributed fairly throughout society \cite{patel2024ethical}.

In this essay, we will examine the intricacies of the AI revolution, its effects on the IT sector, the changing nature of information technology education, and how to steer toward a future in which humans and machines coexist peacefully and work together to create a better world. We'll try to throw light on the opportunities and challenges that lie ahead through thoughtful analysis, well-informed conjecture, and useful suggestions, leading stakeholders on a path of innovation, empowerment, and adaptation in the AI era.

\section{AI Disruption and Its Impact on Work Hours}

AI's unparalleled degree of autonomy has significant ramifications for the nature of work and established employment structures \cite{parker2022automation}. The first industrial revolution led to the establishment of the 40-hour workweek and the five-day workweek in order to meet the demands of manufacturing and production-based economies \cite{yang2021industry}. But businesses, especially those in the IT industry, are seeing a dramatic change in their operational capacities as a result of the introduction of cutting-edge technologies like artificial intelligence \cite{bhatt2022artificial}.

Businesses can produce more and more efficiently thanks to AI, which boosts output and, frequently, generates sizable profits \cite{wamba2020influence}. However, there is a price to this increased efficiency: the possible loss of human labor. The demand for human labor may decline as AI systems grow more advanced and capable of carrying out tasks that have historically been performed by humans. This could result in fewer job opportunities and a reevaluation of the traditional workweek \cite{silver2023ai}.

A few progressive companies have already started looking into different work schedules, such as cutting the typical workweek down from five days to four \cite{exos-4-day-workweek}. Companies that adopt a reduced workweek demonstrate their recognition of the evolving nature of the workforce, as well as the significance of work-life equilibrium and employee welfare in an AI-driven world. Furthermore, a reduced workweek can result in better productivity, happier workers, and less burnout—all important factors to take into account in the disruptive era of artificial intelligence \cite{exos-4-day-workweek}.

Furthermore, companies may find themselves in a better position to adjust to the disruptive powers of artificial intelligence if they reconsider standard work schedules and embrace a reduced workweek. A shortened workweek offers prospects for more jobs to be created in addition to encouraging a better work-life balance. Companies may need to hire more workers to maintain productivity levels if employees are given fewer hours to work, which would lessen the possible impact of AI-driven automation on employment \cite{upreti2021artificial}. 
In addition to ensuring that the advantages of AI technology are distributed more fairly among workers, this move towards a more inclusive and equitable labor market can promote economic resilience \cite{braganza2021productive}. Employers can harness the transformative power of AI while protecting the livelihoods and well-being of their workforce by utilizing creative workforce management strategies like flexible scheduling and remote work options.

\section{Impact on the IT Industry}
The IT industry is well-positioned to capitalize on AI's potential to accelerate digital transformation and create value for clients and stakeholders because it is at the forefront of technological innovation. AI has the potential to completely transform the way IT businesses run and provide services because of its capacity to automate tedious jobs, streamline workflows, and extract valuable insights from massive volumes of data \cite{iansiti2020competing}.

But as AI technologies spread, traditional IT jobs and skill sets are also at risk of becoming extinct. Data entry, analysis, and troubleshooting are among the jobs that AI algorithms and machine learning models are now automating that previously required human intervention. In order to guarantee that their workers stay relevant in an AI-driven world, businesses are being forced by this change to reassess their workforce strategies and make investments in reskilling and upskilling programs.

Furthermore, the rise of AI-powered platforms and services is fostering innovation and collaboration in the IT sector by establishing new ecosystems \cite{iansiti2020competing}. To create and implement cutting-edge AI solutions, businesses are working more and more with AI startups, technology suppliers, and academic institutions. These partnerships are erasing industry boundaries and changing competitive dynamics as established IT companies have to change to stay competitive against AI disruptors and new entrants.

Thereofe, IT companies need to adapt agile, AI-centric operating models in order to stay competitive and meet the changing needs of their clients in this rapidly changing landscape. This involves utilizing technologies to automate repetitive processes, boost productivity, and provide better customer experiences, such as computer vision, natural language processing, and robotic process automation \cite{iansiti2020competing}. IT companies can navigate the challenges and uncertainties of an increasingly AI-driven world while seizing new opportunities for growth, innovation, and value creation by leveraging the power of AI.

\section{ Implications for IT Education}

IT education has two mandates in light of the changing needs of the AI-driven economy: first, to provide students with a foundational understanding and technical proficiency in AI and related fields; second, to encourage critical thinking, creativity, and adaptability in the face of swift technological advancements \cite{xia2022artificial}. To guarantee that graduates have the knowledge and abilities required to succeed in the digital age, a thorough review of curricula, pedagogies, and learning experiences is necessary for the integration of AI into IT education programs.

Including multidisciplinary viewpoints and practical AI applications in the curriculum is one way to reinvent IT education \cite{ahmad2020scenario}\cite{southworth2023developing}. Universities could, for instance, provide interdisciplinary courses that bring together computer science and disciplines like psychology, sociology, criminology, or ethics to give students a comprehensive understanding of how artificial intelligence is affecting society. Furthermore, students can apply their theoretical knowledge in real-world settings through practical projects and experiential learning opportunities like internships or co-ops with industry partners. These opportunities help students develop the problem-solving and teamwork skills necessary for success in the AI-enabled workplace \cite{rovzman2023building}.

Additionally, students' development of ethical reasoning and responsible innovation must be given top priority in IT schools. Students need to be prepared to handle difficult moral conundrums involving bias, privacy, and algorithmic accountability as AI technologies become more commonplace \cite{slimi2023navigating}. The inclusion of courses on AI ethics, governance, and policy in the curriculum guarantees that graduates will have the moral consciousness and judgment required to create AI systems that advance justice and societal well-being \cite{nguyen2023ethical}.

Industry collaborations are essential to closing the knowledge gap between academia and business, giving students access to practical experience and business insights \cite{blankesteijn2021science} \cite{khasawneh2024closing}. Students can gain industry-relevant knowledge and practical skills through collaborative projects, hackathons, and mentorship programs led by professionals in the field. They can also build valuable networks and career pathways \cite{hynes2023educating}.

Furthermore, by encouraging a culture of lifelong learning and continuous skill development, IT education must adjust to the changing nature of work in the AI era \cite{shiohira2021understanding}. The competencies required to thrive in the IT sector are always changing as AI technologies advance quickly \cite{ciarli2021digital}. Because of this, IT schools ought to provide professionals with micro-credentialing opportunities and adaptable, modular learning pathways so they can retrain and upskill in response to changing market demands and emerging technologies \cite{pachler2023unbundling}.

With the aforementioned considerations in mind, IT schools can create the creative pedagogical strategies listed below.

\vspace{1em} 

\textit{Project-Based Learning:} Through project-based learning, students work on real-world assignments or case studies where they must use AI principles and methods to solve challenging issues \cite{morais2021improving}. Students could work together to create AI-driven programs like recommendation engines or chatbots. This method gives participants practical experience in AI development while promoting teamwork, critical thinking, and problem-solving abilities.

\vspace{1em} 

\textit{Flipped Classroom:} With the help of the flipped classroom model, students can learn outside of class at their own pace by accessing instructional materials online. Students apply and reinforce their understanding of AI principles in class through active learning exercises like group discussions and AI coding labs \cite{altemueller2017flipped}. This method raises student engagement and comprehension by making the most of interactive learning and problem-solving time in the classroom.

\vspace{1em} 

\textit{Problem-Based Learning:} Students are given real-world, unstructured problems to solve through problem-based learning, which calls for investigation, analysis, and suggested solutions \cite{tan2021problem}. For instance, students could create AI algorithms to draw conclusions or forecast outcomes by analyzing real-world datasets. This method offers hands-on experience using AI techniques to solve real-world problems while encouraging inquiry, critical thinking, and self-directed learning.

\vspace{1em} 

\textit{Collaborative Learning:} Students that participate in collaborative learning work in small groups to accomplish common learning objectives \cite{smith1996cooperative}. Students in an AI course collaborate on research projects and peer code reviews, among other activities, to benefit from one another's knowledge and viewpoints. Through the development of interpersonal, communication, and teamwork skills, this method fosters a collaborative learning environment where students can exchange ideas and learn from one another.

\vspace{1em} 

\textit{Experiential Learning:} Students can apply their theoretical knowledge in real-world settings through experiential learning, which gives them practical hands-on experience \cite{clark2010potential}. Students obtain industry experience and insights into AI applications through participation in industry-sponsored projects, co-ops, and internships. This strategy provides beneficial exposure to the industry, chances for networking, and insights into the real-world uses of AI technology.

\vspace{1em} 

These pedagogical strategies can be successfully incorporated into IT education programs' curricula to effectively prepare students for success in the AI-driven economy. Through practical assignments, group projects, and real-world experiences, students gain the understanding, abilities, and flexibility required to succeed in a field that is changing quickly.

In conclusion, there are many different ways that the AI era will affect IT education, necessitating careful consideration of pedagogy, curriculum design, and industry collaboration. IT schools can equip graduates to succeed in the dynamic and complex environment of the AI-driven economy by embracing interdisciplinary perspectives, ethical considerations, and experiential learning opportunities.

\section{Conclusion}

As we approach the dawn of the AI era, we must navigate a course that balances the potential transformative power of AI with the preservation of human dignity, equity, and resilience. A new era of technological innovation has begun with the rapid advancements in artificial intelligence, which are reshaping economies, societies, and industries worldwide. Notwithstanding the exhilaration and hope surrounding artificial intelligence, it is imperative to acknowledge and confront the multifaceted obstacles and moral dilemmas that coincide with its extensive implementation.

The IT industry is leading this technological revolution, promoting innovation and having a significant impact on the future of work. 
AI technologies are transforming every facet of the IT landscape, from cybersecurity and data analytics to automation and machine learning. However, this change also means that it is crucial to make sure AI is created and applied in an ethical and responsible way, following the guidelines of openness, accountability, and human-centered design.

In a similar vein, information technology education must adapt to the needs of the AI-driven economy in order to provide students with the knowledge and abilities necessary to succeed in a rapidly shifting digital environment. IT education programs can enable students to become lifelong learners and digital leaders who can navigate the complexities of the AI era by embracing pedagogical approaches that foster creativity, adaptability, and critical thinking.

It is crucial to think about the larger ramifications for the future of work and society in light of the disruption caused by AI. The 40-hour workweek, which was created for the first industrial revolution, might not be appropriate for the AI-enabled economy today. Rethinking work hours and employment practices is becoming more and more necessary as businesses use AI technologies to automate repetitive tasks and streamline operations and guarantee fair distribution of opportunities and resources.

\section{Discussion and Future Directions}

In the discussion and future directions section, we can go into more detail about the potential benefits and challenges of implementing a shortened workweek in response to AI disruption. By reducing the number of workdays from five to four, organizations can provide their employees with greater opportunities for personal development, work-life balance, and leisure time. This shift could lead to increased well-being, morale, and productivity as well as a more wholesome and durable workplace culture. 
The implementation of a shortened workweek does, however, present certain practical difficulties and issues, such as managing workloads, scheduling changes, and preserving operational effectiveness. To properly handle the shift, organizations might need to look into flexible scheduling options, alternative work arrangements, and performance-based rewards.

Furthermore, combating the effects of AI-driven automation on employment calls for a multidimensional strategy that incorporates social policies, financial incentives, and reskilling and upskilling programs. Governments and organizations can enable workers to adapt to changing job requirements and pursue new career pathways in emerging industries by funding lifelong learning initiatives and vocational training opportunities.

Furthermore, it is critical to guarantee that all people have equitable access to the advantages of AI technology, especially underprivileged and marginalized groups. Closing the digital divide and fostering digital inclusion through policies and programs can help close the gap and guarantee that everyone has the chance to engage in the AI-driven economy.

Ultimately, we can build a more equitable and sustainable future for the IT sector and beyond by seizing the opportunities that AI presents and taking proactive measures to address its challenges. The development of talent, socioeconomic empowerment, and ethical AI development should be given top priority if we are to fully utilize AI's potential to improve human capabilities, promote inclusive prosperity, and advance society as a whole.

\bibliographystyle{ieeetr}
\bibliography{references}

\end{document}